# Organic Thermoelectric Textiles for Harvesting Thermal Energy and Powering Electronics

Yuanyuan Zheng[1], Qihao Zhang[2], Wenlong Jin[3], Yuanyuan Jing[1], Xinyi Chen[1], Xue Han[1], Qinye Bao[4], Yanping Liu[1], Xinhou Wang[1], Shiren Wang[5], Yiping Qiu[1,6], Kun Zhang[1*], Chongan Di[3*]

[1]Key Laboratory of Textile Science & Technology (Ministry of Education), College of Textiles, Donghua University, Shanghai 201620, PR China

[2]State Key Laboratory of High Performance Ceramics and Superfine Microstructure, Shanghai Institute of Ceramics, Chinese Academy of Sciences, Shanghai, 200050, China

[3]Beijing National Laboratory for Molecular Sciences, Key Laboratory of Organic Solids, Institute of Chemistry, CAS, Beijing 100190, China

[4]Key Laboratory of Polar Materials and Devices, Ministry of Education, School of Information Science Technology, East China Normal University, 200241, Shanghai, P. R. China

[5]Department of Industrial and Systems Engineering, Texas A&M University, College Station, TX 77843, United States

[6]College of Textiles and Apparel, Quanzhou Normal University, Fujian 362000, China

*Corresponding authors: K.Z. (email: kun.zhang@dhu.edu.cn), and C.D. (email: dicha@iccas.ac.cn)

**Abstract**

Wearable thermoelectric devices show promises to generate electricity in a ubiquitous, unintermittent and noiseless way for on-body applications. Three-dimensional thermoelectric textiles (TETs) outperform other types in smart textiles owing to their out-of-plane thermoelectric generation and good structural conformability with fabrics. Yet, there has been lack of efficient strategies in scalable manufacture of TETs for sustainably powering electronics. Here, we fabricate organic spacer fabric shaped TETs by knitting carbon nanotube yarn based segmented thermoelectric yarn in large scale. Combing finite element analysis with experimental evaluation, we elucidate that the fabric structure significantly influences the power generation. The optimally designed TET with good wearability and stability shows high output power density of 51.5 mW/m$^2$ and high specific power of 173.3 μW/(g·K) at ΔT= 47.5 K. The promising on-body applications of the TET in directly and continuously powering electronics for healthcare and environmental monitoring is fully demonstrated. This work will broaden the research vision and provide new routines for developing high-performance and large-scale TETs toward practical applications.

**Introduction**

Self-powered smart textiles with integrated wearable electronics ($10^0$-$10^1$ mW) are very intriguing for adaptive sensing stimuli (gas, pressure, acoustic, heat and electricity) for healthcare and environmental monitoring[1-3]. The self-powered modules can work based on piezoelectric[4-10], triboelectric[11-17] and thermoelectric principles[18-32] *etc*. Wearable piezoelectric nanogenerators (PENG) and triboelectric nanogenerators (TENG) scavenge mechanical energy for electricity generation with ultrahigh power density of $10^1$-$10^2$ W/cm$^2$ as the external resistance is on the order of mega-ohms (MΩ)[6,13]. They usually involve complex wired circuit (i.e. AC-DC converter), low power density ($10^{-3}$-$10^0$ mW/m$^2$)[33] at small external resistance ($10^1$-$10^3$ Ω which is a common range for real circuits) and intermittent mechanical energy supply, *etc*[4,11]. Flexible solid-state thermoelectric generators (TEGs) without moving parts can convert heat to electricity noiselessly by using the ubiquitous temperature gradient between human body and environment, which is attracting significant attention for on-body applications[34-40].

In contrast to two-dimensional TEGs[38,40-54] performing along in-plane direction, three-dimensional deformable TEGs (3D-TEGs)[34,35,55-62] with vertical aligned TE units can utmostly utilize the body-environment temperature gradient vertical to body skin direction. In contrast to flexible inorganic TEGs[34,36,41,44,51,56,59,60,62,63], the soft and skin-compatible organic thermoelectric textiles (TETs)[35,38,41,46,50,64] are very promising as well for the out-of-plane thermal energy harvesting, owing to their abundant sources, low density, intrinsic flexibility, processability and conformability with fabrics.

Nevertheless, there has been lack of efficient strategies in the scalable manufacture of TETs for sustainably powering wearable electronics for healthcare and environmental monitoring. The main challenges involve: (1) efficient and large-scale fabrication of thermoelectric yarns, (2) manufacturing TETs via industrial textile processes, and (3) high power output for powering electronics in practical applications.

In this work, we address these issues through developing organic-based spacer fabric shaped TETs by using large-scale produced carbon nanotube based segmented thermoelectric yarn (TEY) as spacer yarn. We first find an efficient routine to produce segmented TEY with poly(3,4-ethylenedioxythiophene): poly(styrenesulfonate) (PEDOT: PSS, *p*-type) and polyetherimide (PEI, *n*-type) in large scale, which can generate tens of meters of TEY in minutes. For the first time, the effect of intrinsic fabric structure on the power generation is evaluated with the assistance of finite element simulation and experimental analysis. The spacer fabric shaped TET with good wearability and stability shows high power density of 51.5 mW/m$^2$ and voltage density of 520.9 V/m$^2$ at $\Delta T$= 47.5 K. Furthermore, the TET can directly and continuously power electronics for healthcare and environmental monitoring.

**RESULTS**

**Design and manufacture of segmented TEYs**

We adopt a simple route (Figure 1a) to prepare carbon nanotube based segmented TEY，which can meet the length requirement for industrial textile process. CNT yarn (CNTY, $\phi$~125.8 μm) is wrapped onto a thin polyethylene terephthalate (PET) plate (5 mm wide) and rolled up to a packed cylindrical structure. Subsequently, it is alternately immersed into aqueous PEDOT: PSS and PEI solutions to form the so-called carbon nanotube yarn based segmented TEYs. After fully drying in ambient condition, the TEY is twisted with polyester filament to enhance mechanical stability for further usage. Figure 1b shows a certain part of CNT based segmented TEY before twisting with polyester filament. The length of p-type segment and n-type segment is ~3.5 mm (optical image in Figure 1b). The energy dispersive spectra (EDS) mappings clearly indicate that the segmented TEY possess a periodic distribution of *p*-electrode-*n*-electrode unit along the TEY.

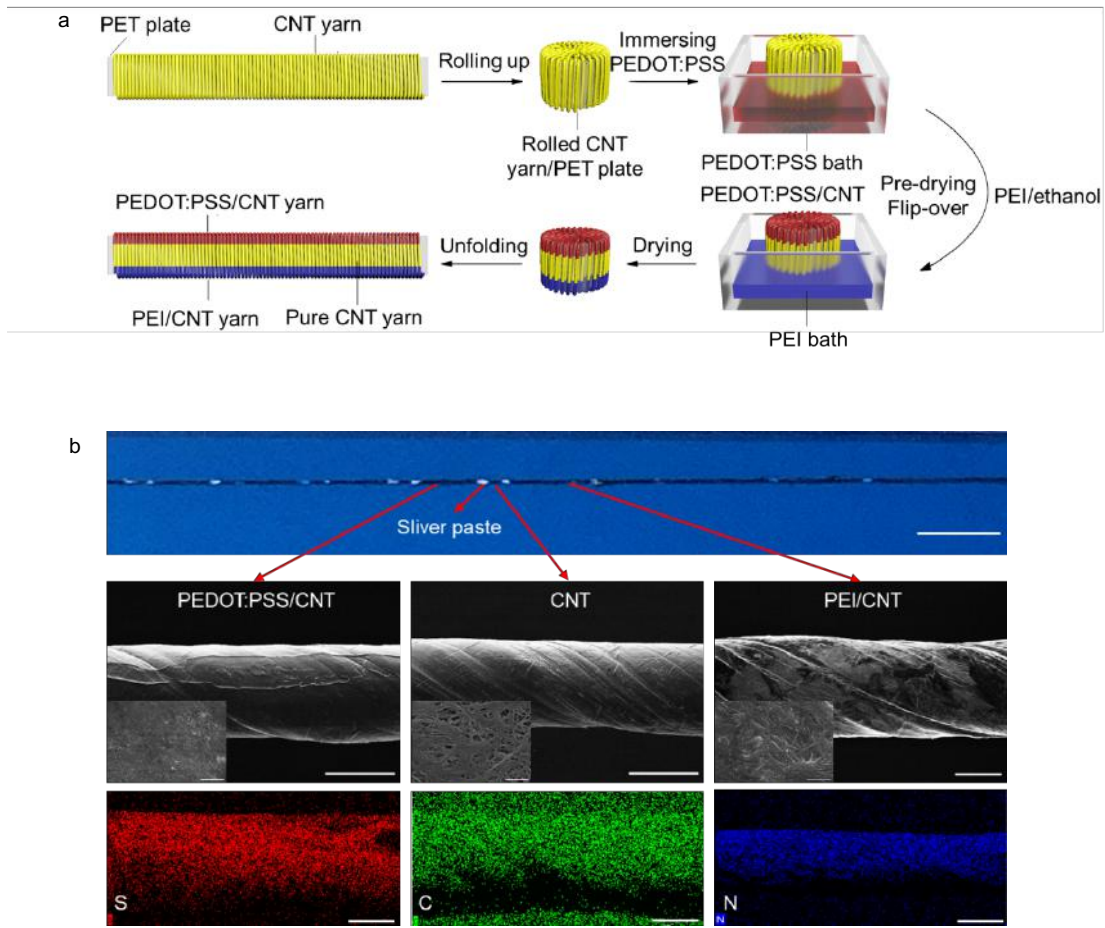

**Figure 1** | **a**. Illustration of the scalable manufacture process of TEYs: Step 1: The CNTY are wrapped on a thin PET plate with a moderate wrapping density; Step 2: The PET plate are rolled for further solution dipping; Step 3: The CNTY/PET roll is soaked in PEDOT:SS solution with a pre-designed dipping length; Step 4: After pre-dried at 40 °C, the CNTY/PET roll is flipped over and immersed into PEI ethanol solution with a certain doping length. After complete drying at ambient condition, the segmented TEY in 10-meters scale has been successfully made for further usage. Since CNTY can wick solutions very quickly, the designed dipping length for PEDOT:PSS and PEI should be shorter than that of designed segment lengths. Step 5: The TEY is twisted with a polyester filament to provide mechanical support and structural stability. **b.** The

optical image (upper figure, scale bar: 5 mm) shows a given section of CNTY based segmented TEY before twisting with polyester filaments. Silver paste is used to mark boundaries between different segments. The scanning electronic microscopy (SEM, middle figures, scale bar: 100 μm) images show a single unit of P-electrode-N junction along TEY. The EDS results (bottom figures, scale bar: 100 μm) distinguish the different segments by mapping their characteristic elements (sulfur S for PEDOT:PSS, carbon C for pure CNT and nitrogen N for PEI).

**Thermoelectric properties of TEYs**

The as-purchased CNTY shows the Seebeck coefficient $S$ of ~54.0 μV/K with the electrical conductivity σ of ~838.1 S/cm (Supplementary Table 1) leading to the average power factor of 246.1 μW/(m·$K^2$). The PEDOT:PSS/CNT composite yarn shows average $S$ of 70.1 μV/K, $σ$ of 1043.5 S/cm and power factor of 512.8 μW/(m·$K^2$) (Supplementary Table 2), which is almost 2-folds of that of original CNTYs. We selected commonly used nitrogen-abundant polymer PEI to tune the thermoelectric properties of CNTY. In Figure 2a, the PEI doping can finely tune the thermoelectric performance of CNTY form p-type to n-type. The optimized PEI/CNTY possesses an average negative $S$ of -68.7 μV/K, $σ$ of 1408.3 S/cm and power factor of 667.8 μW/(m·$K^2$) (Supplementary Table 3), suggesting the efficient n-doping of PEI to CNTY. Here, we confirm the PEI n-type doping with the assistance of UPS and UV-Visible spectra (Figure 2b, 2c and Supplementary Figure 1) by identifying the band structure change in doped carbon nanotubes.

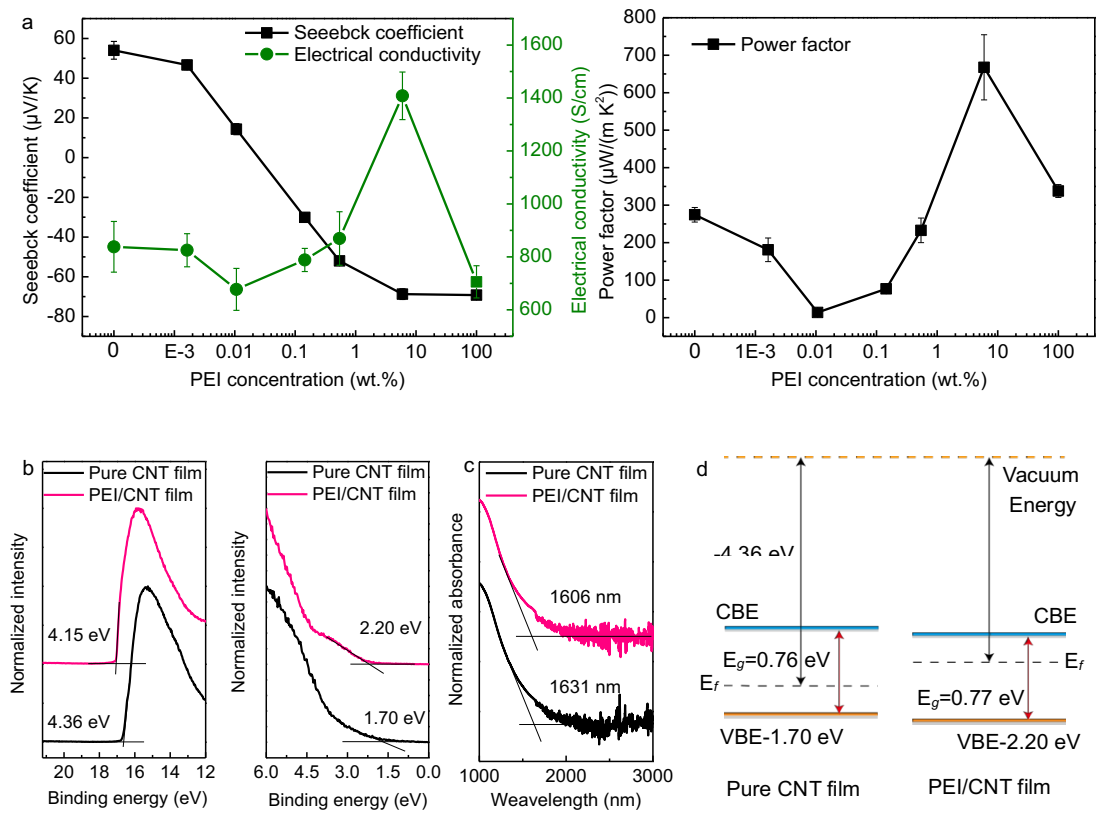

**Figure 2** | **a.** The thermoelectric properties of CNTY as a function of PEI concentrations in ethanol. **b.** UPS and **c.** UV-vis spectra of pure CNT film and PEI doped CNT film. **d.** The band diagrams of original CNT and PEI-doped CNT according to the measured UPS and UV-Vis spectra. Note: The CNT film and PEI-doped CNT film share the same CNT source with TEYs.

Figure 2d shows the band structures of the original CNT and PEI-doped CNT (Supplementary Note 1) according to the measured UPS and UV-Visible spectra of these samples. The work function (~4.15 eV) of PEI-CNT is smaller than that of original CNT (~4.36 eV) while the band gap is almost the same, which confirms the n-type doping of PEI. This might be due to ethylamine molecular and ethylamine/CNT interface dipole moments[65] after the physical sorption of ethylamine groups on CNT surfaces. More interestingly, the PEI doping can switch the $S$ sign of CNTY from positive to negative with much higher $|S|$ ($|S|$=68.7 μV/K) and $σ$ (1408.3 S/cm) values than those of original CNTY. This should be related to the abundant feature of density of states (DOS) near the Fermi level of CNT after PEI doping since only the DOS in a few $kT$ ($k$ is the Boltzmann's constant; T is the absolute temperature) contributes to thermoelectric transport.

**The effect of textile structure on the thermoelectric generation of TETs**

Before fabricating and investigating the thermoelectric power generation performance of CNTY-based TETs, we firstly study the textile structural effect on the thermoelectric performance of TETs because different textile structures have different thermal properties that would lead to large variation of the final power generation performance. Three typical fabric structured TETs (Figure 3a-3c), which are featured with stacked knitted fabric, warp-knitted spacer fabric and stacked woven fabric are constructed based on the Solidworks software. With the assistance of finite element analysis (FEA), the generation performances of segmented TEYs (single *p-n* pair is simulated for simplification) are fully investigated. Details in the simulation are described in Supplementary Figure 2, Supplementary Table 5 and Supplementary Table 6. Here, a fixed temperature is applied onto the bottom surface (hot-side) of the modules while the top surface (cold-side) of the modules is set as contacting with ambient air by thermal convection. At low temperature (<100 ºC), thermal convection and conduction are dominated in heat transfer within textiles, and the thermal radiation can be neglected.

Taking the hot-side temperature of 35 °C as an example, the temperature and potential distributions within TETs are shown in Figure 3a-3c.

In order to clearly describe the temperature distribution along TEYs, we simply hide the textile substrates (the second and third rows in Figure 3a-3c). The third row displays the potentials generated in the TEYs embedded in different fabric configurations. The built temperature difference along the TEY embedded in knitted fabric is 5.79 K, which is much larger than that of spacer fabric (4.48 K), woven fabric (3.04 K) and bared single *p-n* pair of TEY (1.89 K, Supplementary Figure 3 and Supplementary Figure 4). The corresponding potentials are 0.699 mV, 0.559 mV, 0.406 mV and 0.248 mV for weft-knitted fabric, spacer fabric, woven fabric shaped TETs and bared single *p-n* pair of TEY (Supplementary Figure 3 and Supplementary Figure 4), respectively. Overall, the knitted fabric results in higher thermal resistance in the thickness direction. As we know that woven fabrics usually have regular yarn arrangement in thickness direction, hence there are fewer closed channels (Figure 3c) that can localize static air to suppress thermal convection through woven fabric. Though the knit loops are periodically connected with each other, weft-knitted fabric (Figure 3a) or warp-knitted spacer fabric (Figure 3b) usually have irregular yarn arrangement overall, thus they may localize more static air for reduced thermal convection in these fabrics. We alter the hot-side temperature to further verify this trend. In Figure 3d, the woven fabric shaped TET shows the smallest slope (the ratio of output voltage to temperature) at a given hot-side temperature, while the knitted fabric shaped TET outperforms other fabric structured ones.

Based on the simulation results, we experimentally prepare three corresponding TETs (nine pairs of p-n pair of TEY) with a fixed thickness (~3.5 mm) (a fixed TEY percentage in TET) by stacking several layers for knitted (6 layers, Supplementary Table 7) or woven fabrics (22 layers, Supplementary Table 7) because it is difficult to fabricate thick knitted or woven fabrics. The specific weight of spacer fabric, knitted fabric and woven fabric are 346.9 $g/m^2$, 659.1 $g/m^2$, 2465.8 $g/m^2$, respectively (Supplementary Table 7). We then measure the built-in temperature difference and output voltage to corroborate the simulation results. (Note: Like the simulation, we do not apply heat sink on the top surfaces of fabrics.) Though the specific fabric weight for three TET samples is very challenging to control, the measured results are still helpful for understanding the effect of fabric structure on the power generation. In Figure 3e, the stacked knitted fabric shaped TET has the highest output voltage. Moreover, the spacer fabric shaped TET shows slightly higher voltage, because the specific fabric weight (2465.8 $g/m^2$, Supplementary Table 7) for stacked woven fabric is much higher than that of spacer fabric (346.9 $g/m^2$, Supplementary Table 7). So, it is reasonable to believe that the thermal resistance in this stacked woven fabric is higher than that of spacer fabric. In addition, we also investigate the structural parameters of spacer fabric shaped TETs including density and thickness (Supplementary Figure 5) in warp-knitted spacer fabric. The results show that with the increasing of thickness, the output voltage increases as well. This is because when the thickness is larger there will be more static air trapped in the spacer fabric, thus leading to a lower thermal conductivity of the whole TET. It is much easier to fabricate spacer fabrics with a large thickness (up to 30 cm) during industrial textile process. Hence, to demonstrate the real thermoelectric generation performance of TETs, we finally select spacer fabric as the target fabric structure.

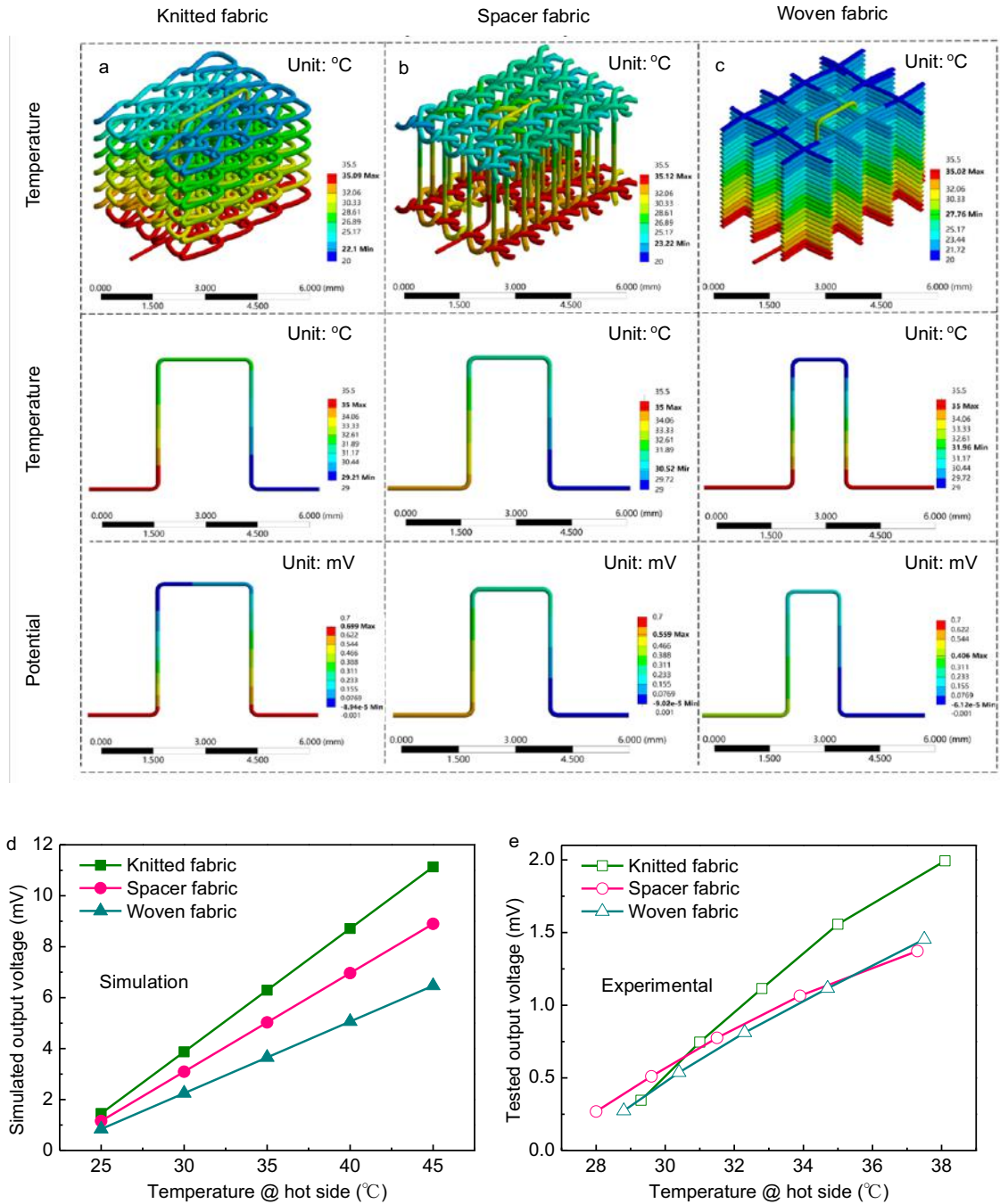

**Figure 3 | a-c.** Finite element simulation results of three typical fabric structured TETs including weft-knitted, warp-knitted spacer fabric and woven at a given specific fabric

weight and fixed hot-side temperature (35 °C). The first row shows the temperature distribution for whole TETs. To clearly see the temperature and potential distributions of TEYs in the TETs. We hide the textile substrate in the second and third rows. The second row shows the temperature distribution of TEY by hiding textile substrate. The third row shows the potential distribution of TEY by hiding textile substrate. **d.** The simulated open-circuit voltage of three TET modules. Note: The voltage value is multiplied by nine for comparing with the measured TETs (nine pairs of p-n pair of TEY). **e.** The measured open-circuit voltage of three TETs. Note: We prepared three different types of TET by fixing the thickness (~3.5 mm). The thickness of knitted and woven fabrics is achieved to 3.5 mm by stacking several layers, which will lead to largely different specific fabric weights. The voltage was measured without any heat sink on the TETs.

**Thermoelectric power generation of TET**

With comprehensive consideration of the fabric structure effect and the difficulty in practical fabrication, we prepare a warp-knitted spacer fabric shaped TET (lateral size: 8 cm×9.3 cm, 3.5 mm in thickness) with 966 pairs of *p-n* pairs (Figure 4a and 4b). The lateral areal density of *p-n* pairs in spacer fabric is ~7.7 $mm^2$ per pair. The TET shows excellent flexibility (Figure 4c, 4d and 4e), signifying the good conformability with human skins. It can be bended, twisted and compressed arbitrarily without deteriorating the thermoelectric performance (Supplementary Figure 6).

The output voltage and power density of spacer fabric-based TET under various ΔT are shown in Figure 4f and Figure 4g, respectively. As a given ΔT is applied along the out-of-plane direction of TET, the output voltage is linear to the applied ΔT, showing voltage density up to 520.9 V/m² at ΔT=47.5 K (Figure 4f). According to the power output formula $P = U^2R/(r + R)^2$, maximum power density $P_{max}$ occurs as internal resistance $r$ is equal to the external $R$. $P_{max}$ of TET reaches up to 51.5 mW/m² (382.8 µW, Figure 4g) at ΔT=47.5 K when $R$=9.50 kΩ. This has been the highest values ever reported for the organic TETs and outperforms most of wearable TEGs (organic and inorganic based ones) [22, 33, 36, 38-40]. In Figure 4h, the short circuit current is negatively linear to the output voltage, achieving a short circuit current of ~402 µA at ΔT=47.5 K, which stands in a considerably high level to power some typical IoT electronics. Higher output voltage and power can be further realized by preparing TET in larger size with higher areal density of *p-n* pairs using more segmented TEYs. In Figure 4i, we normalize the reported thermoelectric power outputs and term as "specific power" by dividing them by the mass of thermoelectric materials and the applied temperature gradient in unit of µW/(g·K). This normalization is reasonable to evaluate the capability of power generation for all types of wearable TEGs. Because different kinds of TEGs have their own structural configurations and may involve different accessories, it is challenging to assess the device performance in aspects of device size or mass. Our TET shows a specific power of 173.3 µW/(g·K) which is much higher than other types of wearable TEGs, approaching to that of $Bi_2Te_3$ based TET (250.9 µW/(g·K))[60]. The high power output performance is believed to originate from the proper selection of fabric structure and careful optimization of thermoelectric properties of TEYs. Additionally, the thermo-wet comfortability of TET is not obviously affected by the insertion of TEYs compared with the as-received warp-knitted spacer fabric (Supplementary Note 2 and Supplementary Table 8), which indicates that the TET has an obvious advantage in wearable electronics.

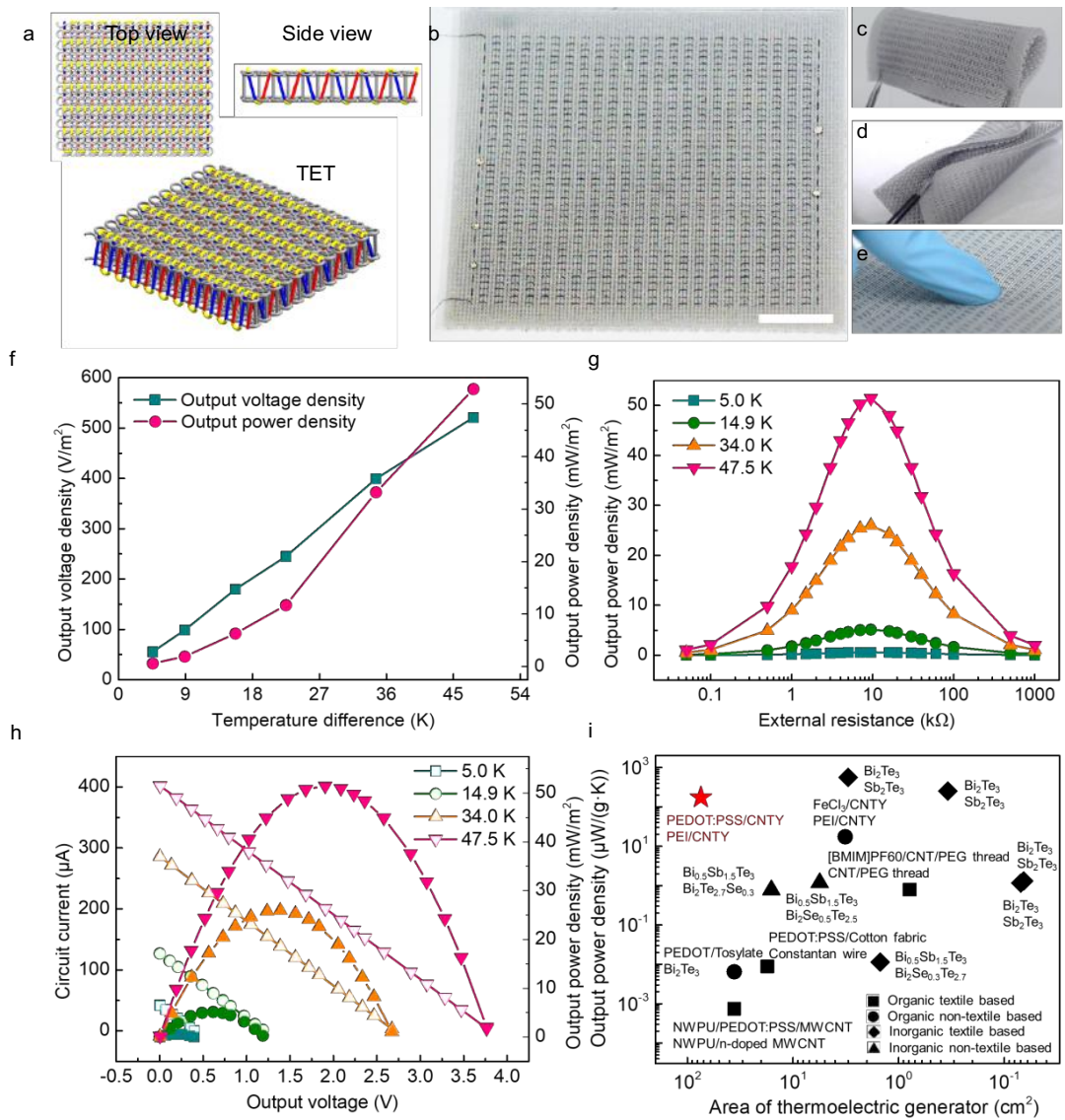

**Figure 4** | Thermoelectric power generation performance of warp-knitted spacer fabric shaped TET. **a-e.** Diagrams (**a**) and optical images of as-prepared warp-knitted TET. The TET sample (**b**) is 8 cm×9.3 cm in lateral size (scale bar: 2cm), showing excellent

flexibility. The illustrated flexibility of TET including bending (**c**), twisting (**d**) and compressing (**e**). **f.** Output voltage density and power density of TET with respect to $\Delta T$. Note: The temperature difference $\Delta T$ for measuring output voltage and power was applied by direct heating the bottom side of TET without any heat sinks on the top surface. **g.** Output power density of TET as a function of external resistance R at different $\Delta T$. **h.** Circuit current and power density as a function of voltage output at different $\Delta T$. **i.** The comparison of normalized power density of TET in our work with the data reported in literature including textile-based and non-textile based TEGs (Supplementary Table 9).

**Practical applications of TET for powering wearable electronics**

Power sustainability is imperative for real-time healthcare and environmental monitoring. To realize energy autonomy, a reliable power supply capable of generating electricity unintermittingly in extreme weathers is urgently demanded. Yet, there have been few reports on the successful practical demonstrations of organic-based TETs that can directly power and/or charge commercial electronics due to their limited electricity output. We demonstrate that our TET can power several typical electronics including body thermometer (Yuwell, YT318, 1.5V/30μW, Figure 5b, Supplementary Video 2), thermo-hygrometer (MITRI, 1.5V/30μW, Figure 5c, Supplementary Video 2), ultraviolet detector (MEET, MS-98(3), 3V/30μW, Figure 5d, Supplementary Video 2), pedometer (Meilen, mj001, 3V/75μW, Figure 5e, Supplementary Video 2) and electronic watch (Zengda, jsqdzb001, 3V/6.5μW, Supplementary Video 2) at $\Delta T\sim20$-

45 K for healthcare and environmental monitoring (Figure 5a and 5f). Because the output voltage of our TET can be steady at ~3.85 V as $\Delta T$ ~47.5 K (Figure 4h) with a current of ~402 μA. To well understand the stability of TET power output, we first charge a supercapacitor (KAMCAP, SE-5R5-Z104UY, 0.1F) with an initial potential of 0V to 1.5-3 V as required by various electronics and subsequently use the supercapacitor to power electronics. As shown in Figure 5b-5e, the charging time is much smaller than that of discharging time, indicating the TET can continuously and stably generate enough electricity to power electronics. Figure 5f shows the scenarios of direct powering electronics by TETs.

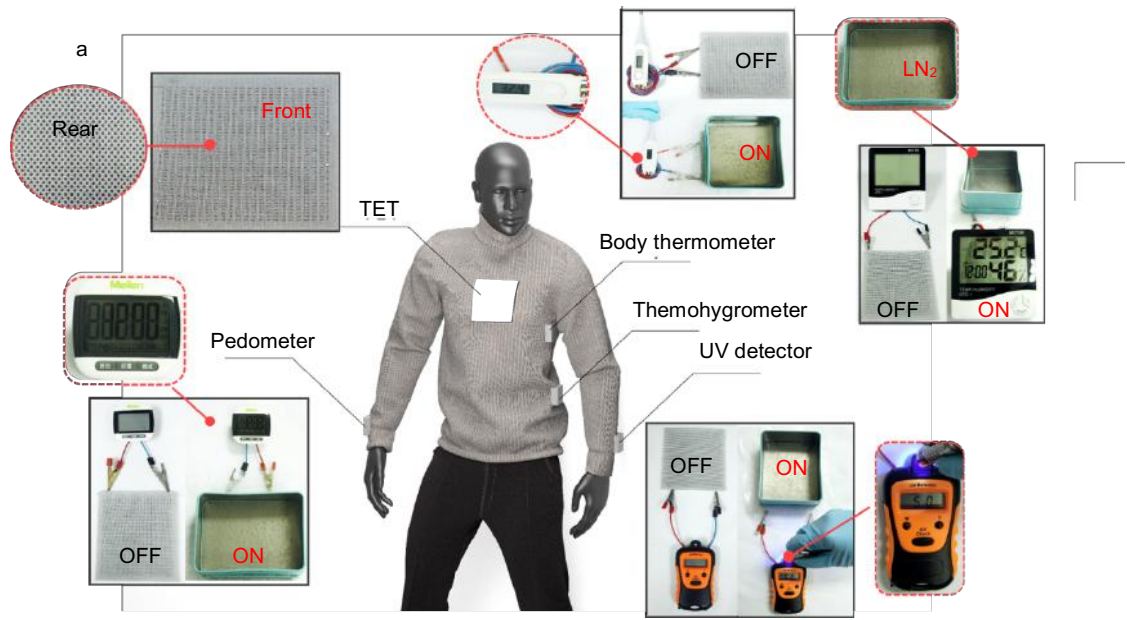

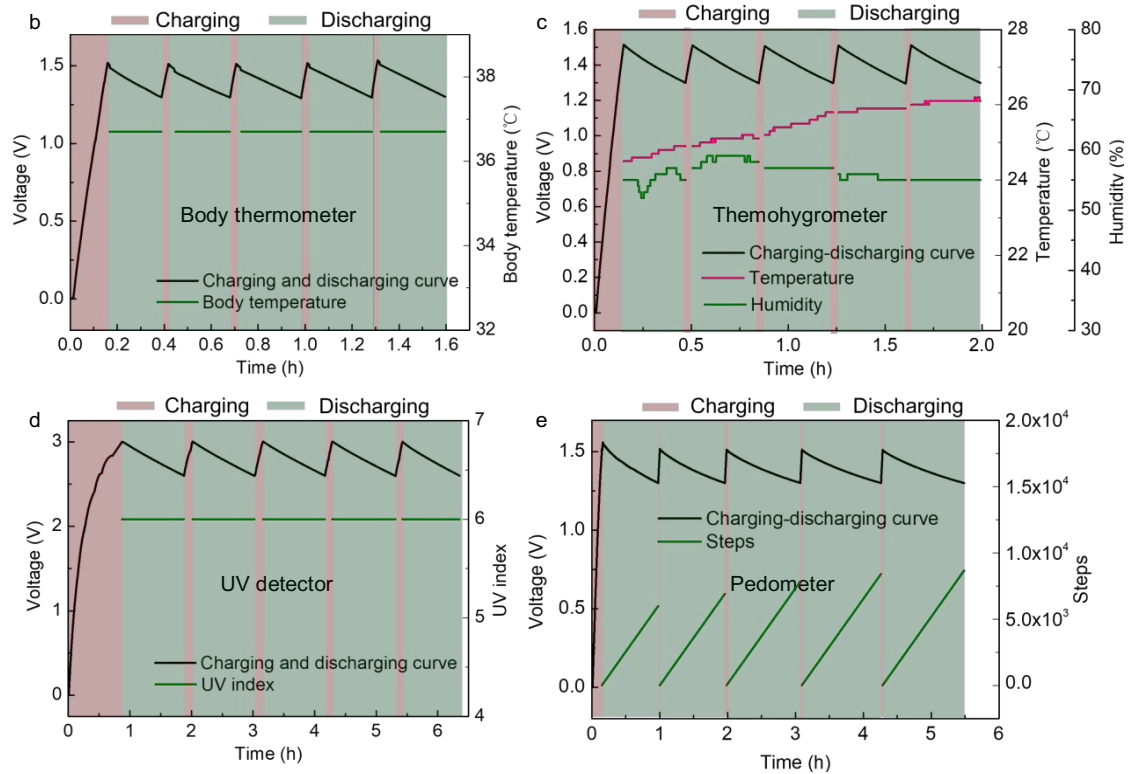

**Figure 5** | **a.** Illustration of an energy-autonomy clothing integrated with TET and wearable sensor network. **b-e.** Body thermometer **(b)**, thermo-hygrometer **(c)**, ultraviolet detector **(d)** and pedometer **(e)** powered by the spacer fabric shaped TET. f. Photos of TET direct powering electronics.

**Scalable manufacture of spacer fabric shaped TETs**

To further prove the industrial applications of our TEY and TETs, we adopt a commercialized computerized flat knitting machine to directly knit TEYs into a weft-knitted spacer fabric. The TEYs twisted with polyester filament are further braided with a polyester fabric layer using a braiding machine to provide additional anti-friction capability (Figure 6a, Supplementary Video 3). Weft-knitted spacer fabrics consist of two separate outer fabric layers joined together but kept apart by spacer yarns. The spacer fabric was knitted on a computerized flat knitting machine Stoll ADF 530-32W (Reutlingen, Germany) with a gauge of E7.2. The needles installed in the needle beds were E10. The yarn path diagram of this 3D structure is shown in Supplementary Figure 7. The outer layers were knitted with two nylon/spandex covered yarns (70D/20D) using single jersey structure, while the spacer yarn was tucked to the two outer layers with a polyester monofilament of a diameter 0.12 mm. The TEY was also knitted as a spacer yarn to connect the two outer layers and form the final product. Detailed knitting process and material selections were summarized in Supplementary Note 3 and Video 4. Overall, this section signifies the scalable manufacturing of TET using conventional industrial textile process.

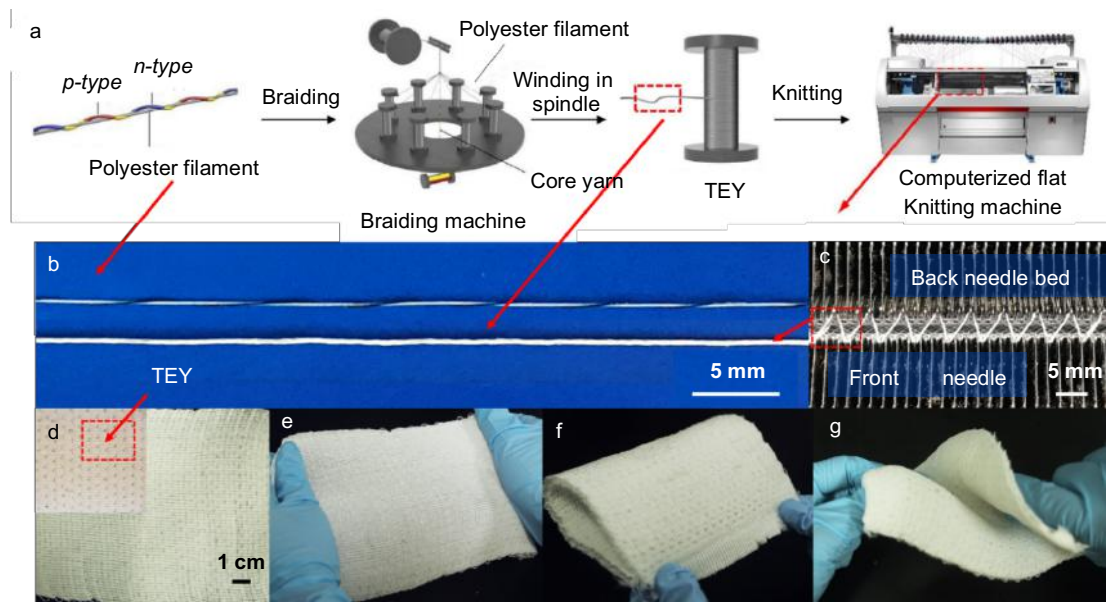

**Figure 6 |** Demonstrating scalable manufacture of weft-knitted spacer fabric based TETs. **a.** Illustration of industrial weft-knitting process with a computerized flat knitting machine. The twisted core yarn comprised of TEYs with polyester filament are further braided with a fabric sheath of polyester filaments with improved anti-friction property for high-speed textile processes. A commercialized computerized flat knitting machine (Reutlingen, Germany) was used to fabricate the TET. **b.** The optical images showing the twisted TEY with thick polyester filament and core-sheath TEY. **c.** The cross-section view of braided TEYs in spacer fabric during knitting. The TEY forms a zigzag configuration in the cross-section of spacer fabric. **d.** The as-prepared spacer fabric with TEYs. The dark spots are TEY, showing uniform distribution of TEY in spacer fabric. **e-g.** The flexibility of spacer fabric shaped TET under stretching (**e**), bending (**f**) and twisting (**g**).

**Discussion**

In summary, we develop a strategy to prepare segmented thermoelectric yarn using carbon nanotube yarn and scale up the manufacture of thermoelectric textile using matured industrial textile process. We explore the effect of fabric structural configuration on the power generation of thermoelectric textiles, the fundamental understanding is instructive for designing thermoelectric textiles. A spacer fabric shaped thermoelectric textile with high power density of 51.2 mW/m$^2$ at a temperature difference of 47.5 K is achieved and its promising on-body applications in powering electronics is fully demonstrated.

**Methods**

**Materials.** PEDOT: PSS solutions (CLEVIOS PH1000), polyetherimide (PEI, Aladdin, M. W. 600, 99%), dimethyl sulphoxide (DMSO, Sinopharm) were all used as received. The CNTY were purchased from the Nano Institute of Suzhou and used as purchased.

**Morphology characterizations.** The microscopic morphology and EDS element analysis were characterized by FE-SEM (Hitachi, SU8010 series) using an accelerating voltage of 5.0 kV (morphology) and 10.0 kV (EDS) and work distance of ~8 mm (morphology) and ~15 mm (EDS).

**FEA simulation of TET**

We construct all models using Solidworks (Version 2014). All the meshing processes are done in Ansys Workbench (Version 18.0).

**The fabrication process of TET.** The segmented TEY was wrapped on a polyester filament and then braided by 8 rolls of polyester yarn using a 2D-circular knitting machine (KBL-16-2-90, Xuzhou Henghui Braiding Machine CO., LTD). This core-sheath segmented CNTY was knitted into a weft-spacer fabric using a computerized flat knitting machine Stoll ADF 530-32W (Reutlingen, Germany).

**Thermoelectric property measurements of TEYs.** The room-temperature thermoelectric measurements (Seebeck coefficient, electrical conductivity, open-circuit voltage, short-circuit current and maximum output power) of TEYs were conducted using home-made thermoelectric measurement systems. Seebeck coefficient was calculated by S=-$\Delta$V/$\Delta$T, where the thermo-voltage was measured by Keithley 2182A and the temperature difference was calculated with the assistance of thermal infrared camera (Fotric 226). The electrical conductivity of all TEYs was measured by 4-wire method using Keithley 2400. The diameter of samples was detected by an optical microscope (Nikon ECLIPSE LV100POL). Each sample was measured about 50 points and average values were used in the manuscript.

**Thermoelectric power generation performance of TET.** The different temperature difference is generated putting TETs between a temperature self-controlled hot plate and metal heat sink. The output power was calculated by the equation $P = U^2R/(r + R)^2$, $U$ is the open circuit voltage, $r$ and $R$ are the internal resistance of TET and applied external resistance. $R$ is altered with a variable resistor.

**Demonstration of practical applications of TET for electronics.** The TET first charged a commercial supercapacitor (5.5V, 0.1F) by applying a temperature difference with liquid nitrogen without any heat sinks, and then powered electronics with sufficiently high voltage and current. The TET was also used tp directly power electronics for proving its capability.

**Data availability**

The data supporting our findings in this work are available from the corresponding author K.Z. upon reasonable request.

organic electronics. *Science* **336**, 327-332 (2012).

**Author Contributions**

K.Z. conceived the idea, designed the experiment. K.Z. and C.D. guided the project. Y.Z. performed fabric modeling. Y. Z and Q. Z. performed finite element simulation and data analysis. Y.Z. prepared all TEY and TET samples. Y. Z. and J.W. conducted thermoelectric property measurements of yarns and data analysis with C.D.. Y.Z. measured the power generations of TETs and analyzed data. Y.J. partially performed the fabric manufacturing. Y. J., X.C. and X.H. conducted device demonstrations, drew all figures and made all videos. Y.L. guided the weft-knitted fabric manufacturing. Q.B. performed UPS and UV-Vis measurements and data analysis. K.Z., Y.Z, Q.Z., C.D., W.J. prepared the initial manuscript. All authors discussed the results and contributed to review the manuscript.

**Acknowledgement**

K. Z. thanks the financial support from the Fundamental Research Fund for the Central Universities (19D110106), the National Natural Science Foundation of China (No.51603036), Young Elite Scientists Sponsorship Program by CAST (2017QNRC001), and the "DHU Distinguished Young Professor Program". We acknowledge Dr. Fujun Xu at College of Textiles, Donghua University for providing raw materials at early stage in this project. We acknowledge Dr. Lidong Chen at Shanghai Institute of Ceramics, Chinese Academy of Sciences for giving us valuable advices. We also acknowledge Ms. Xinzhi Hu graduated from College of Textiles, Donghua University for facilitating us to do the pre-experiment at early stage in this project.

*Supplementary Information for*

**Organic Thermoelectric Textiles for Harvesting Thermal Energy and Powering Electronics**

Yuanyuan Zheng[1], Qihao Zhang[2], Wenlong Jin[3], Yuanyuan Jing[1], Xinyi Chen[1], Xue Han[1], Qinye Bao[4], Yanping Liu[1], Xinhou Wang[1], Shiren Wang[5], Yiping Qiu[1,6], Kun Zhang[1*], Chongan Di[3*]

[1]Key Laboratory of Textile Science & Technology (Ministry of Education), College of Textiles, Donghua University, Shanghai 201620, PR China

[2]State Key Laboratory of High Performance Ceramics and Superfine Microstructure, Shanghai Institute of Ceramics, Chinese Academy of Sciences, Shanghai, 200050, China

[3]Beijing National Laboratory for Molecular Sciences, Key Laboratory of Organic Solids, Institute of Chemistry, CAS, Beijing 100190, China

[4]Key Laboratory of Polar Materials and Devices, Ministry of Education, School of Information Science Technology, East China Normal University, 200241, Shanghai, P. R. China

[5]Department of Industrial and Systems Engineering, Texas A&M University, College Station, TX 77843, United States

[6]College of Textiles and Apparel, Quanzhou Normal University, Fujian 362000, China

[*]Corresponding authors: K.Z. (email: kun.zhang@dhu.edu.cn), and C.D. (email:

dicha@iccas.ac.cn)

**Contents**

**Supplementary Note 1** The preparation of pure CNT film and PEI doped CNT film.

**Supplementary Note 2** Thermo-wet comfortability of TET

**Supplementary Note 3** Design parameters and knitting process of weft-knitted spacer fabric shaped TET using TEYs.

**Supplementary Figure 1 |** The full spectra of **a.** UPS and **b.** UV-vis of pure CNT film and PEI/CNT film.

**Supplementary Figure 2 |** Mesh details of TET with different substrate structures. **a.** Warp-knitted spacer fabric shaped TET. **b.** Knitted fabric shaped TET. **c.** Woven fabric shaped TET. **d.** Bared single *p-n* pair of TEY.

**Supplementary Figure 3 |** The FEA simulation results of bared single *p-n* pair of TEY at fixed heating temperature (35 °C). **a.** The temperature distribution of the bared single *p-n* pair of TEY. **b.** The potential distribution of bared single *p-n* pair of TEY.

**Supplementary Figure 4 |** The simulated thermoelectric output voltage of bared single *p-n* pair of TEY.

**Supplementary Figure 5 |** The effect of fabric thickness in spacer fabric on the thermoelectric power generation of TET.

**Supplementary Figure 6 |** Stability of **a.** Seebeck coefficient and **b.** electrical conductivity of TET under twisting, bending, compression and stretching.

**Supplementary Figure 7 |** Lapping diagram of TEY and textile yarn.

**Supplementary Table 1** Thermoelectric properties (electrical conductivity($\sigma$), Seebeck coefficient ($S$) and power factor (PF)) of as-purchased CNTY.

**Supplementary Table 2** Thermoelectric properties ($\sigma$, $S$ and PF) of PEDOT: PSS/CNTY.

**Supplementary Table 3** Thermoelectric properties ($\sigma$, $S$ and PF) of PEI/CNTY.

**Supplementary Table 4** The work function (WF) and valence band edge (VBE) of pure CNT and PEI doped CNT film.

**Supplementary Table 5** Input engineering data in Workbench.

**Supplementary Table 6** Analysis settings in Ansys Workbench.

**Supplementary Table 7** The specific fabric weights of three types of fabrics used for TETs.

**Supplementary Table 8** The thermo-wet comfortability of the as-received warped-knitted spacer fabric and warped-knitted spacer fabric shaped TET.

**Supplementary Table 9** The detailed data and calculating process of Figure 4i.

**Supplementary References**

## Supplementary Notes

### Supplementary Note 1

**The preparation of pure CNT film and PEI doped CNT film**

The pure CNT films were purchased from Suzhou Xiyin nanotechnology company. The pure CNT films were spin coated on a clean indium tin oxide (ITO) glass substrate with a size of 10 mm×6 mm×~50 nm in which the kind of multi-walled carbon nanotubes (MWCNTs) is the same as the MWCNTS in CNTY. For the PEI-CNT film, the concentration of PEI in ethanol is 0.8 mM. Each PEI-CNT film contains 1 ul PEI/ethanol solution.

### Supplementary Note 2

**Thermo-wet comfortability of TET**

The air Permeability was measured by YG461H fully automatic permeability instrument (fabricated by NINGBO textile instrument factory) according to GB/T 11048-2008.

Supplementary Table 7 shows that all measured parameters referring to the thermo-wet comfort are very similar to the as-received spacer fabric and TET, indicating little effect of incorporated TEY on the wearing performance. The air permeability of TET is 10% lower than that of the space fabric. The thermal resistance and Clo value of TET are increased by 9.7% as compared with that of untreated warp-knitted spacer fabric. After threading the TEY into the spacer fabric, the heat transfer coefficient is 91.1% of that of spacer fabric, indicating a better thermal insulation performance.

**Supplementary Note 3**

**Design parameters and knitting process of weft-knitted spacer fabric shaped TET using TEYs.**

Needle setting：66；

Yarn：nylon/spandex covered yarns×2 (70D/20D)，polyester monofilament（0.12 mm）。

1. Technical Parameters:

Machine：ADF 530-32W，Stitch length 7.2，Needle hook gauge 10；

Stitch length：Top and bottom layer plain 11.5，polyester monofilament tuck 8.6，investigated yarn tuck 9；

Speed：0.5 m/s，(Speed is slowed down by manual control during actual knitting)

fabric take-up：1.7。

During the knitting process, the yarn were periodically fed into the computerized flat knitting machine (ADF 530-32 W, Germany Stoll) as spacer yarns to form weft-knitted spacer fabrics using two sets of needles. Weft-knitted spacer fabrics consist of two separate outer knitted layers interconnected by spacer yarns. Producing spacer fabrics on flat machines is creating two independent fabric layers on the front- and back-needle beds separately and then connecting them by tucks on both the needle beds. The distance between the two needle beds determines the spacer fabric thickness, which can be finely tuned to match the CNT yarn based thermoelectric legs. The specific knitting process is as follows: first, the yarns are knitted on the front and back needle bed to form a row of needles with the top outer layer and the bottom outer layer, respectively. Then the yarn is knitted on the front and back needle bed at the same time. Finally, spacer fabrics with 14 needle spacing are knitted on the machine.

We notify that the knitting fabric shows relatively higher thermoelectric output power as the fabric weight and height are the same. Honestly, it is much easier to fabricate thick spacer fabrics (up to 30cm thick) than other types of textiles. Moreover, it is clearly to monitor the moving track of CNT yarn during knitting process.

**Supplementary Figures**

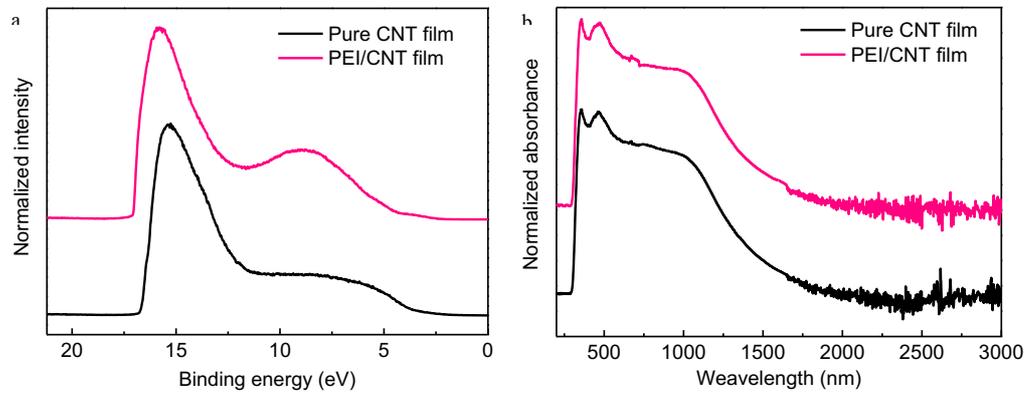

**Supplementary Figure 1** | The full spectra of **a.** UPS and **b.** UV-Vis of pure CNT film and PEI/CNT film.

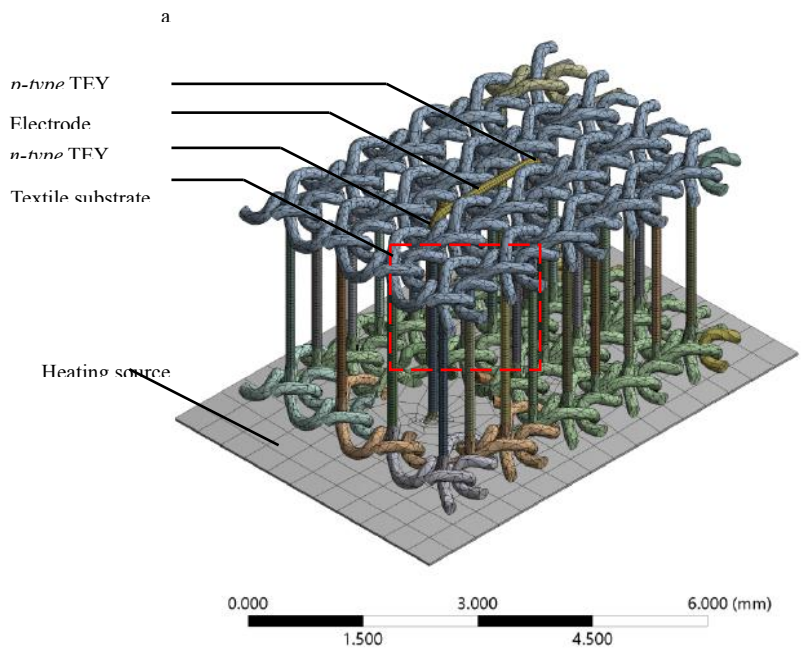

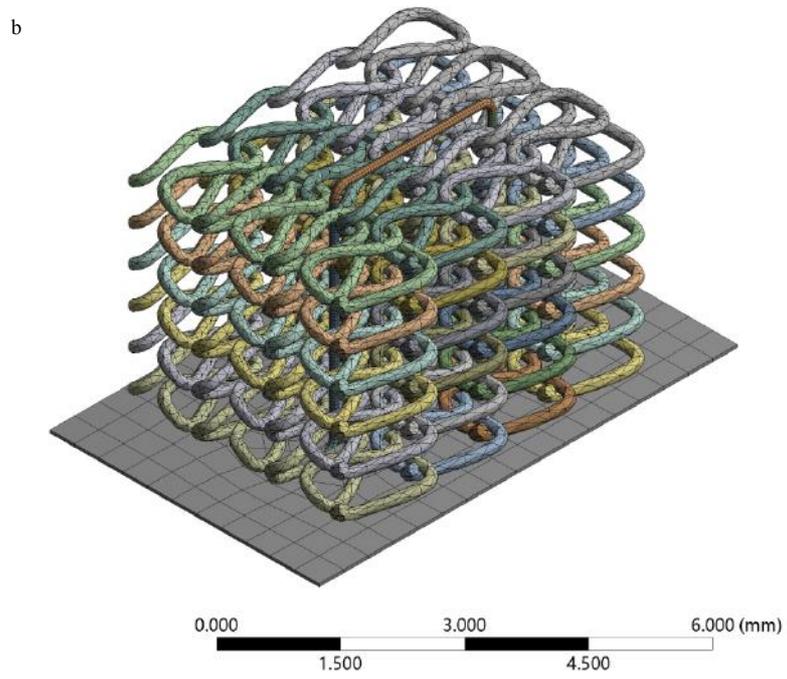

c

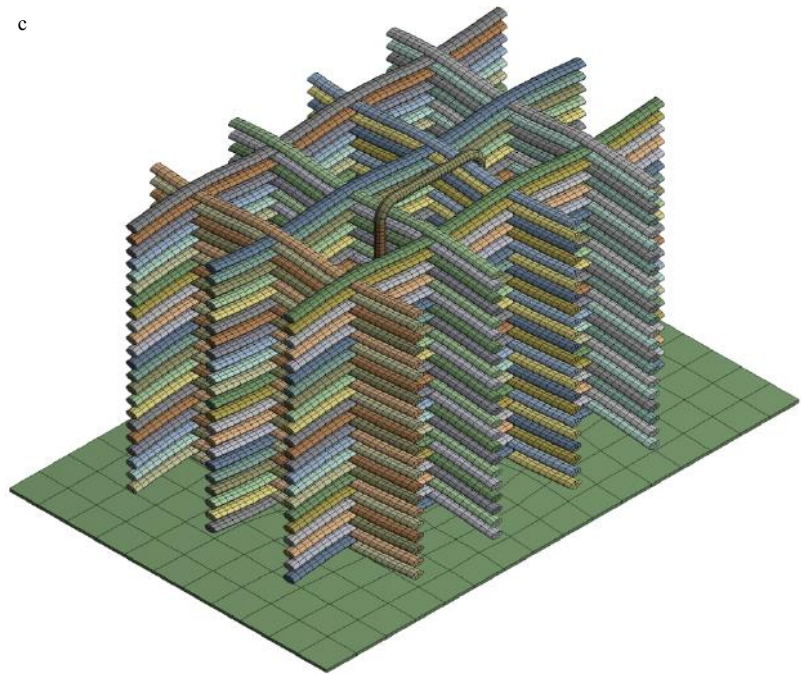

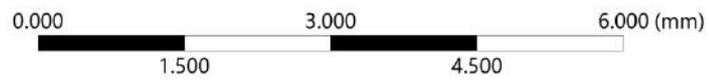

d

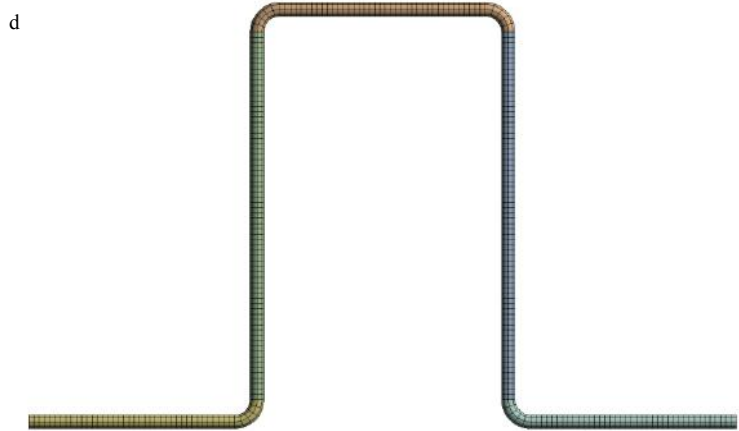

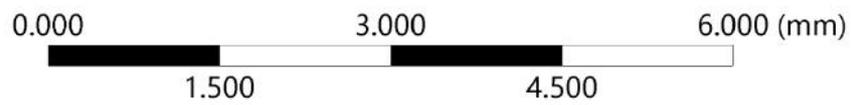

**Supplementary Figure 2 |** Meshing details of TET with different fabric structures. **a.** Warp-knitted spacer fabric shaped TET. **b.** Stacked knitted fabric shaped TET. **c.** Stacked woven fabric shaped TET. **d.** Bared single *p-n* pair of TEY.

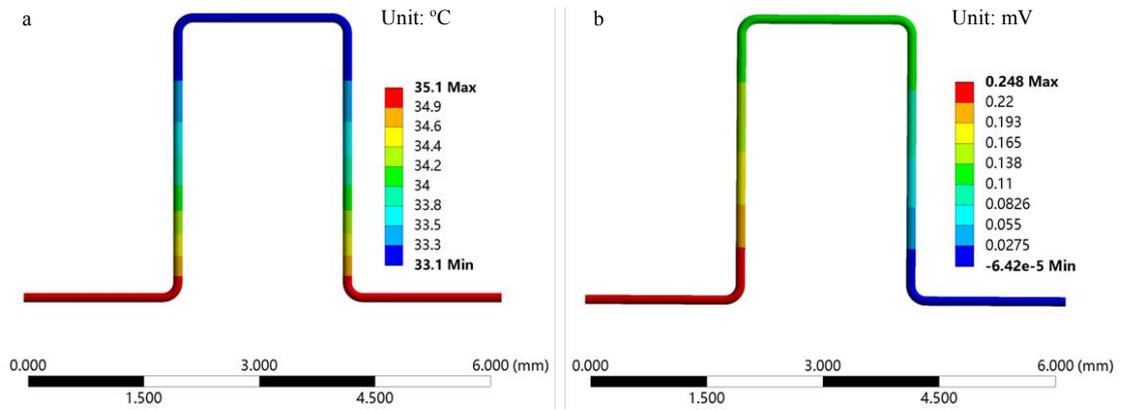

**Supplementary Figure 3 |** The FEA simulation results of bared single *p-n* pair of TEY at fixed heating temperature (35 °C). **a.** The temperature distribution of the bared single *p-n* pair of TEY. **b.** The potential distribution of bared single *p-n* pair of TEY.

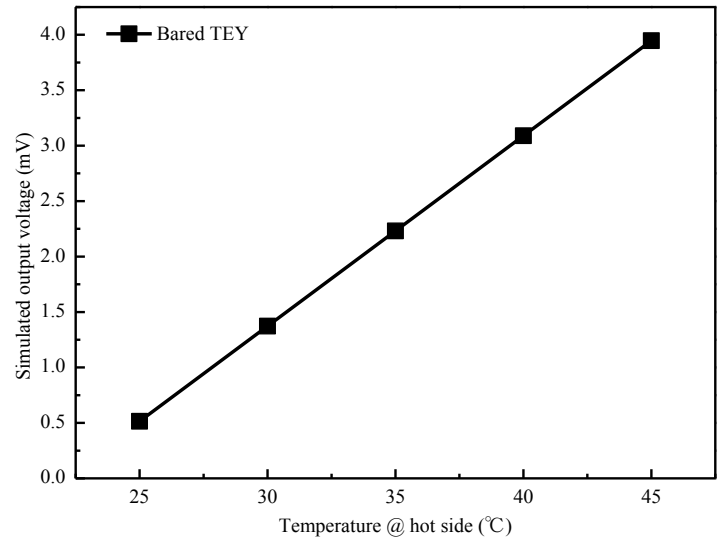

**Supplementary Figure 4** | The simulated thermoelectric output voltage of bared single *p-n* pair of TEY.

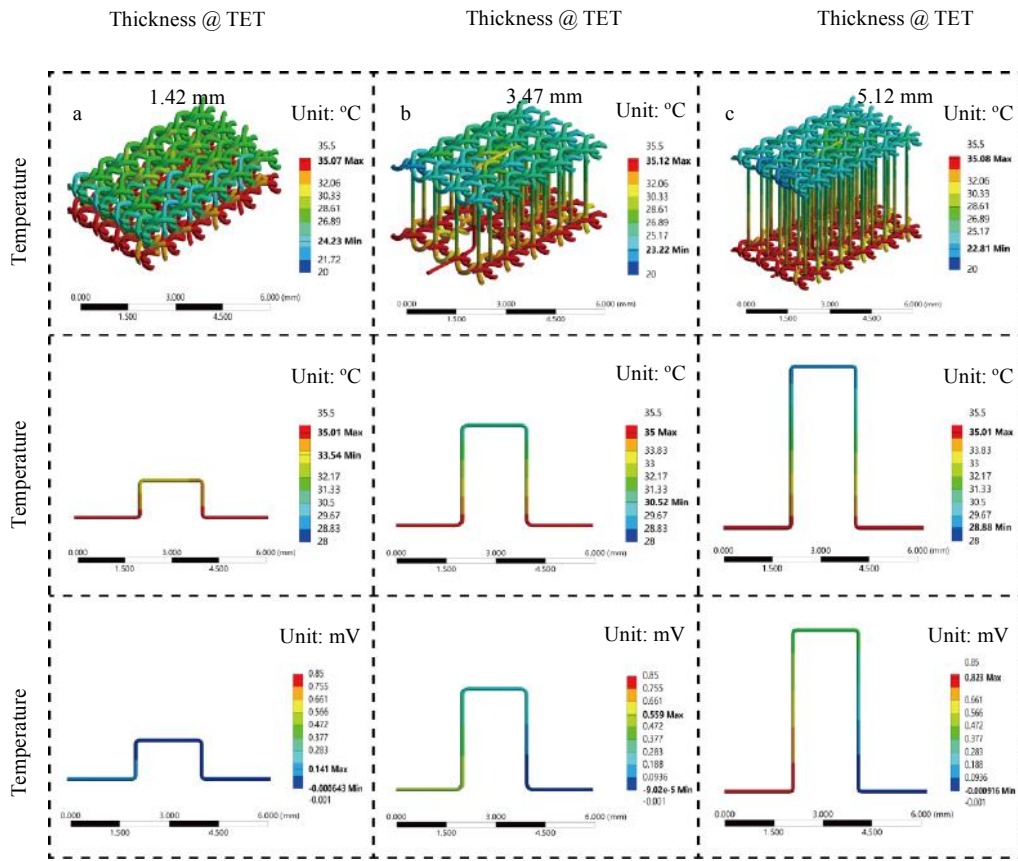

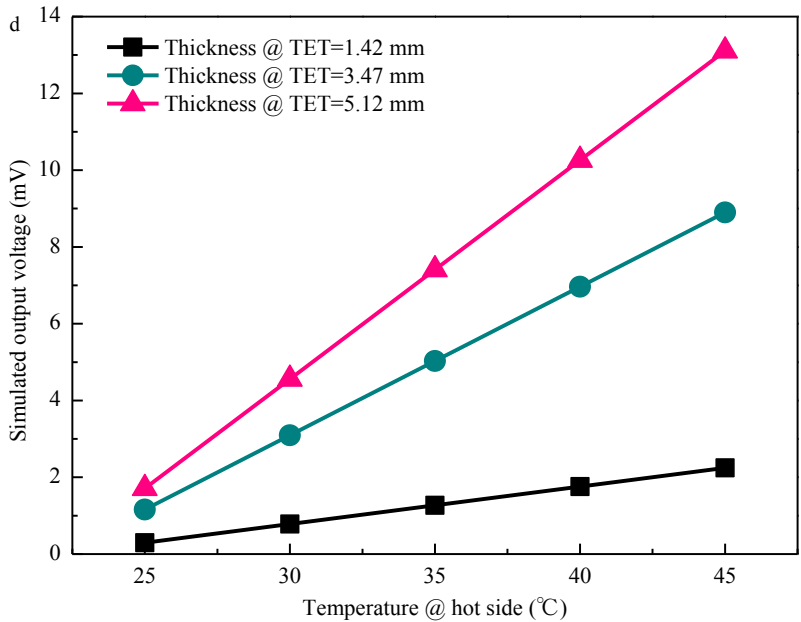

**Supplementary Figure 5 |** The effect of fabric thickness in spacer fabric on the thermoelectric power generation of TET.

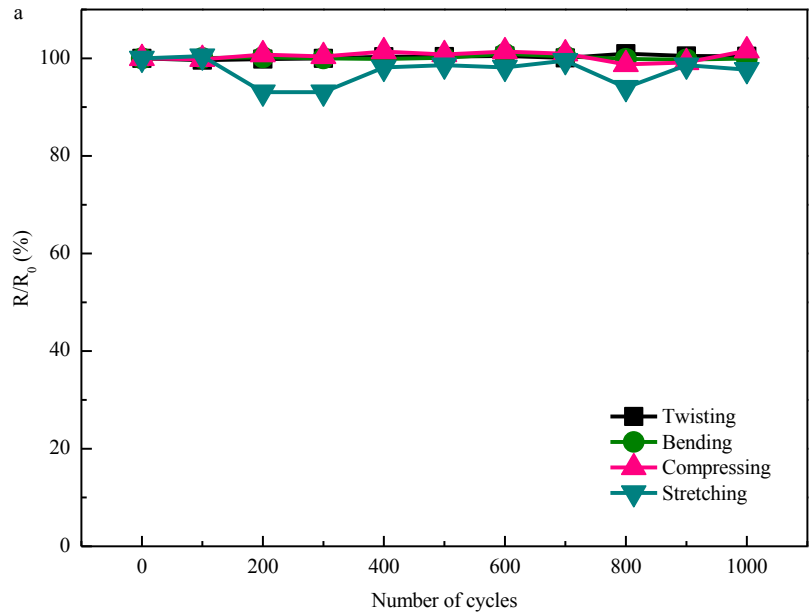

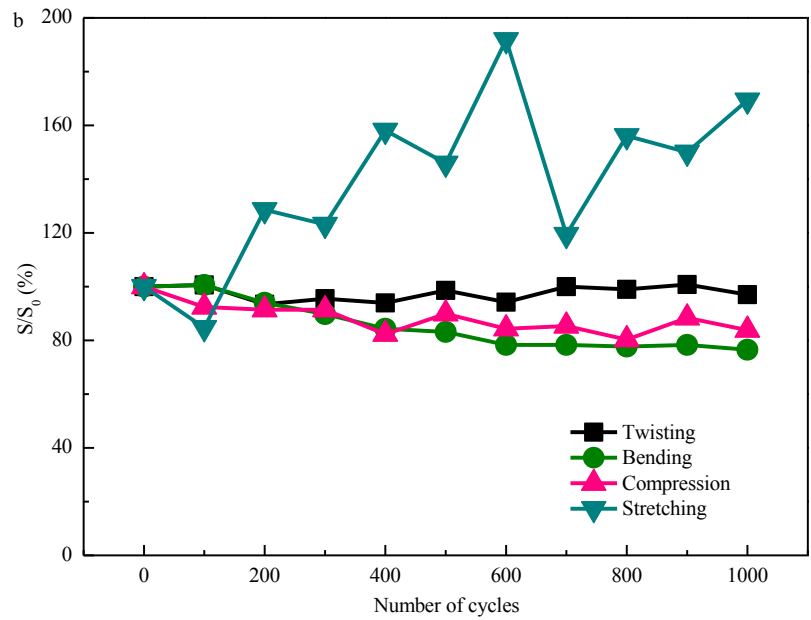

**Supplementary Figure 6 |** Stability of **a.** Seebeck coefficient and **b.** electrical conductivity of TET under twisting, bending, compression and stretching.

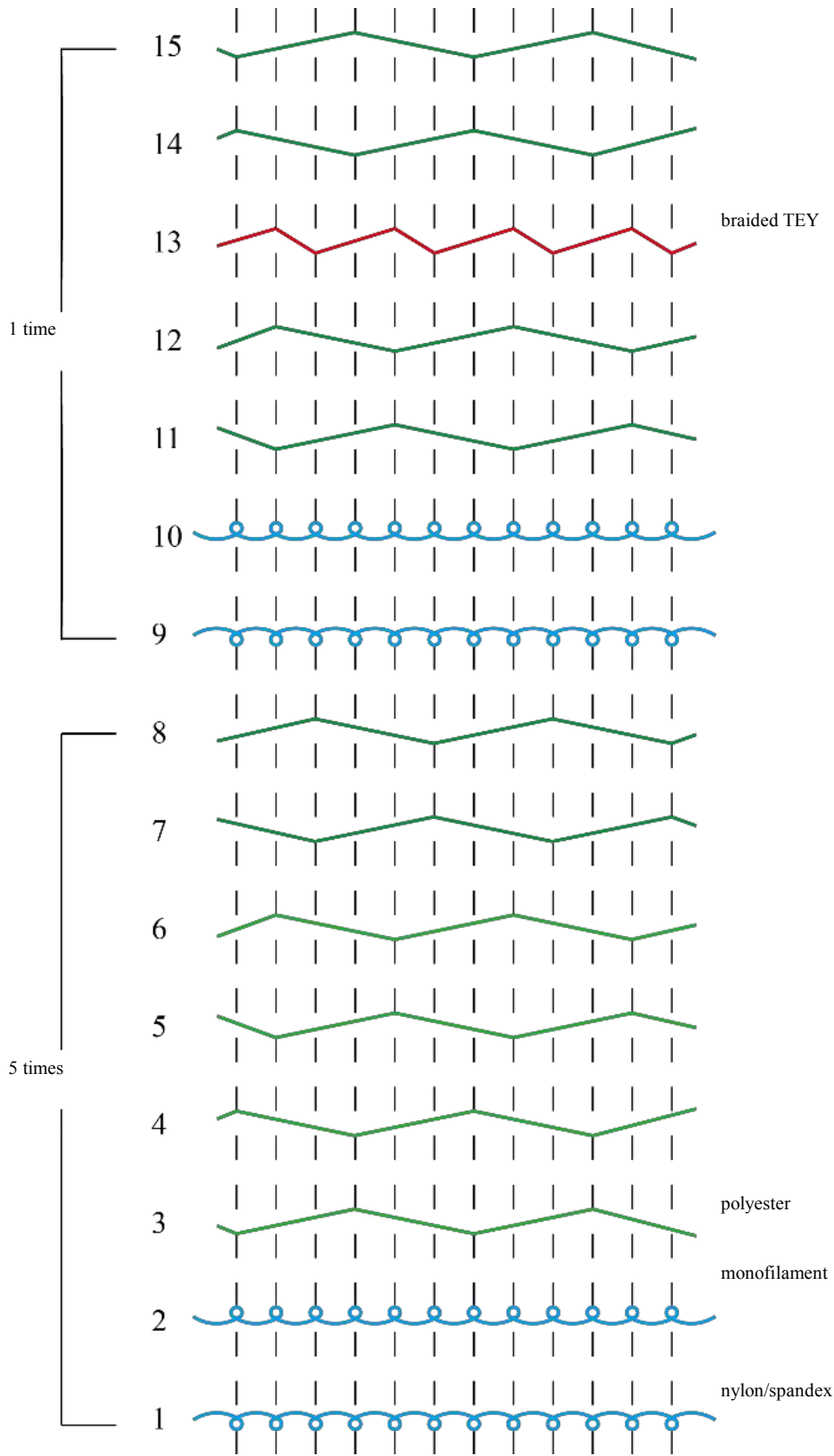

**Supplementary Figure 7** | Lapping diagram of TEY and textile yarn.

**Supplementary Tables**

**Supplementary Table 1** Thermoelectric properties (electrical conductivity($\sigma$), Seebeck coefficient ($S$) and power factor (PF)) of as-purchased CNTY.

| Sample | Resistance ($\Omega$) | Length (cm) | Diameter ($\mu$m) | $\sigma$ (S/cm) | $S$ ($\mu$V/K) | PF $\mu$W/(m·K$^2$) |
|---|---|---|---|---|---|---|
| As-purchased CNTY | 4.49 | 0.50 | 122.99 | 938.23 | 55.80 | 292.13 |
| | 4.93 | 0.50 | 125.68 | 817.75 | 55.68 | 253.52 |
| | 5.08 | 0.50 | 126.40 | 785.17 | 46.26 | 168.02 |
| | 5.08 | 0.50 | 115.99 | 931.83 | 54.58 | 277.59 |
| | 4.78 | 0.50 | 136.29 | 717.48 | 57.72 | 239.04 |
| Average | | | | 838.09 | 54.01 | 246.06 |

**Supplementary Table 2** Supplementary Table 2 Thermoelectric properties (σ, *S* and PF) of PEDOT: PSS/CNTY.

| Sample | Resistance (Ω) | Length (cm) | Diameter (μm) | σ (S/cm) | *S* (μV/K) | PF μW/(m·K²) |
|---|---|---|---|---|---|---|
| PEDOT: PSS/CNTY | 3.81 | 0.50 | 126.31 | 1047.28 | 69.37 | 503.97 |
| | 4.09 | 0.50 | 118.62 | 1107.04 | 68.84 | 524.62 |
| | 4.03 | 0.50 | 131.79 | 909.11 | 69.95 | 444.83 |
| | 4.13 | 0.50 | 117.83 | 1110.54 | 72.14 | 577.94 |
| Average | | | | 1043.49 | 70.08 | 512.84 |

**Supplementary Table 3** Thermoelectric properties (σ, *S* and PF) of PEI/CNTY.

| Sample | Resistance (Ω) | Length (cm) | Diameter (μm) | σ (S/cm) | *S* (μV/K) | PF μW/(m·K²) |
|---|---|---|---|---|---|---|
| PEI/ CNTY | 3.66 | 0.50 | 115.42 | 1304.34 | -68.64 | 614.53 |
| | 3.43 | 0.50 | 117.75 | 1339.14 | -66.78 | 597.20 |
| | 3.70 | 0.50 | 110.25 | 1414.32 | -69.24 | 678.05 |
| | 3.12 | 0.50 | 115.50 | 1529.10 | -72.94 | 813.52 |
| | 3.40 | 0.50 | 113.39 | 1454.54 | -66.10 | 635.52 |
| Average | | | | 1408.29 | -68.74 | 667.77 |

**Supplementary Table 4** The work function (WF) and valence band edge (VBE) of pure CNT and PEI doped CNT film.

|  | WF (eV) | VBE (eV) |
| --- | --- | --- |
| Pure CNT film | 4.36 | 1.70 |
| PE/CNT film | 4.15 | 2.20 |

**Supplementary Table 5** Details regarding material's parameters used in the finite element analysis.

| Materials | Thermal conductivity (W/(m·K)) | Electrical resistivity (ohm·cm) | Seebeck coefficient (μV/K) |
|---|---|---|---|
| Electrode (CNTY) | 35.00 | 0.0011 | / |
| p-type TEY | 35.00 | 0.00096 | 70.08 |
| n-type TEY | 35.00 | 0.00073 | -68.59 |
| Insulating layer | 0.20 | / | / |
| Textile substrate | 0.97 | / | / |

Note: We measured the thermal conductivity of CNTY by a self-heating 3ϖ method.[1] The thermal conductivities of p-type and n-type TEYs are estimated based on that of CNTY. The thermal conductivity of textile substrate is based on the reported thermal conductivity of polyester filament.

**Supplementary Table 6** Boundary conditions applied in the finite element analysis.

| Condition | Materials | Value |
|---|---|---|
| Temperature | Heating source | 25-40 °C |
| Convection 1 | Outer layer of textile substrate | 5 W/(m²·°C) |
| Convection 2 | Inner layer of textile substrate | 0.5 W/(m²·°C) for the spacer fabric; 0.1 W/(m²·°C) for the knitted and |

|         | woven fabric |
| ------- | ------------ |
| Voltage | A surface of electrode | 0 V |

Supplementary Table 7 The specific fabric weights of three types of fabrics used for TETs.

| TET structure | Single layer fabric | | Stacked fabric | |
| --- | --- | --- | --- | --- |
| | Specific weight (g/m$^2$) | Thickness (mm) | Specific weight (g/m$^2$) | Thickness (mm) |
| Spacer fabric | 346.93 | 3.50 | 346.93 | 3.50 |
| Knitted fabric | 109.85 | 0.58 | 659.10 | 3.50 |
| Woven fabric | 112.08 | 0.16 | 2465.76 | 3.50 |

**Supplementary Table 8** The thermo-wet comfortability of the as-received warped-knitted spacer fabric and warped-knitted spacer fabric shaped TET.

| Sample | Air Permeability (mm/s) | Thermal resistance (cm$^2$·K/W) | Heat transfer coefficient | Clo | Insulation rate (%) |
|---|---|---|---|---|---|
| As-received warped-knitted spacer fabric | 2894.00 | 83.52 | 11.97 | 538.80 | 58.36 |
| Warp-knitted spacer fabric based TET | 2603.00 | 91.62 | 10.91 | 591.10 | 60.59 |

**Supplementary Table 9**

**The detailed data and calculate process of Fig.4i**

Table. Detailed data in References

| Size of TEG (cm×cm) | Weight of TE materials (g) | Area of TEG (cm²) | Output power (μW) | Output power per unit weight (μW/g) | Temperature difference (K) | Output power density (μW/g K) | TEG type | Ref |
|---|---|---|---|---|---|---|---|---|
| / | 17.44*a1 | 16.00 | 78.00 | 4.472 | 5.75*a2 | 0.78 | Inorganic non-textile based | 2 |
| 4.30×1.30 | 0.56*b1 | 5.59 | 23.00 | 41.367 | 35.00 | 1.18 | Inorganic non-textile based | 3 |
| 2.50×0.60 | 1.31*c1 | 1.50 | 0.22 | 0.170 | 15.00 | 0.011 | Inorganic textile based | 4 |
| 53.50*d1 | 0.32*d2 | 17.50 | 0.21 | 0.655 | 74.30 | 0.0088 | Organic textile based | 5 |
| 6.00×6.00*e1 | 2.25*e2 | 36.00 | 0.12 | 0.051 | 7.90 | 0.0065 | Organic non-textile based | 6 |

| | | | | | |
|---|---|---|---|---|---|
| / | 0.0099*[f1] | 0.066*[f2] | 0.72 [f3] | 55.00 | Inorganic textile-based | 7 |
| / | 0.0099*[g1] | 0.070*[g2] | 0.63 [g3] | 55.00 | Inorganic textile-based | 8 |
| / | 0.0015*[h1] | 0.34*[h2] | 21.21*[h3] | 55.00 | Inorganic textile-based | 9 |
| 6.00×6.00 | 0.053*[i1] | 36.00 | 13799.610 | 66.00 | Organic textile based | 10 |
| 1.02×0.77*[i1] | 0.00013*[i2] | 0.79 | 0.0492 | 5.00 | Organic textile based | 11 |
| 1.50×2.00 | / | 3.00 | 0.00050 | 50.00 | Inorganic textile-based | 11 |
| 8.00×0.40 | / | 3.20 | / | 17.43 | Organic non-textile based | 12 |
| 8.00×9.30 | 0.048 | 74.40 | 8268.14 | 47.70 | Organic textile based | This work |

Note: Calculate process of the table.

*[a1] Calculated data: Ref 2 presents the weight of p-type TE material is 1.02 g and the weight of n-type TE materials is 1.16 g. There are 8 thermocouples in this TEG, so the weight of the whole TE materials is 1.12×8+1.16×8=17.44 g.

*[a2] Estimated data: The data is obtained from Fig. 7b in Ref 2.

*[b1] The density of p-type $Bi_{0.5}Sb_{1.5}Te_3$ is about 6.8 g/cm$^3$, the size of the p-type $Bi_{0.5}Sb_{1.5}Te_3$ is 1.15 mm×1.15 mm×1.2 mm, so the weight of p-type $Bi_{0.5}Sb_{1.5}Te_3$ is about 1.15×1.15×1.2÷1000×6.8=0.01079 g. The density of n-type $Bi_2Se_{0.5}Te_{2.5}$ is about 7.8 g/cm3, the size of the p-type $Bi_2Se_{0.5}Te_{2.5}$ is 1.15 mm×1.15 mm×1.2 mm, so the weight of p-type $Bi_{0.5}Sb_{1.5}Te_3$ is about 1.15×1.15×1.2÷1000×7.8=0.01238 g. There are 24 thermocouples in this TEG, so the weight of the whole TE materials is 0.01079×24+0.01238×24=0.56 g.

*[c1] Estimated data: The diameter of the p-type and n-type TE materials is 4 mm and the height is 0.6 mm, so the volume is about π×2$^2$×0.6=7.54 mm$^3$=0.00754 cm$^3$. The density of p-type $Bi_{0.5}Sb_{1.5}Te_3$ is about 6.8 g/cm3, so the weight of p-type $Bi_{0.5}Sb_{1.5}Te_3$ is about 0.00754×6.8=0.0512 g. The density of n-type $Bi_2Se_{0.3}Te_{2.7}$ is about 7.73 g/cm3, so the weight of n-type $Bi_2Se_{0.3}Te_{2.7}$ is about 0.00754×7.73=0.0583 g. So the total weight of TE materials is 0.0512×12+0.0583×12=1.31 g

*[d1] Estimated data: The data is obtained from Fig. 3b in Ref 5.

*[d2] Estimated data: The size of each cotton TE strip is 35 mm×5 mm, the area density of the cotton TE strip is assumed to be 0.037 g/cm$^2$. There are 5 strips, so the total weight of the TE legs is about 3.5·0.5·0.037×5=0.32 g.

*[e1] Estimated data: The data is obtained from Fig. 1b in Ref .

*[e2] Estimated data: The size of the p-type film is 18 mm×2 mm×600 nm and the density of the film is assumed to be 1.01 g/cm$^3$. So the weight of the p-type film is about 1.8×0.2×600×10$^{-7}$×1.01=2.1816×10$^{-5}$ g. The size of the n-type $Bi_2Te_3$ is 2 mm×2 mm×2 mm and the density of it is about 7.8 g/cm$^3$. So the weight of n-type leg is 0.2×0.2×0.2·7.8=0.0624 g. There are 36 thermocouples, so the total weight of TE materials is about 7.8 g/cm$^3$.

*f1 Estimated data: The area density of the n- or p-type sheets was about 0.15 mg/cm², the size of each PAN sheet can be estimated to be 4.5 cm×1.22 cm from Fig. 2a. There are 6 PAN sheets in this fabric, so the total weight of the TEG is 0.15×2×4.5×1.22÷1000×6=0.0099 g.

*f2 Calculated data: Ref 6 presents the output power of per couple is 0.24 μW and the power density is 0.11 W/m². There are 3 TE couples. So the area of TEG is 0.24×3÷(0.11×10⁶)×10⁴=0.066cm².

*f3 Estimated data: Ref 6 presents the output power of per couple is 0.24 μW. There are 3 TE couples. So the output power of the TEG is 0.24×3=0.72μW.

*g1 Estimated data: This is the same as data*f1.

*g2 Calculated data: Ref 7 presents the output power of per couple is 0.21 μW and the power density is 0.09 W/m². There are 3 TE couples. So the area of TEG is 0.21×3÷(0.09×10⁶)×10⁴=0.070 cm².

*g3 Estimated data: Ref 7 presents the output power of per couple is 0.21 μW. There are 3 TE couples. So the output power of the TEG is 0.21×3=0.63 μW.

*h1 Estimated data: The length of each p/n segment in the TE twisted yarn can be estimated to be 1 mm from Fig. 2b in Ref 8. There are 21 thermocouples in this plain-weave fabric. So the total length of the TE part is 1×21×2=42 mm. The width of the PAN sheet is about 1.22 cm. The area density of the n- or p-type sheets was about 0.15 mg/cm². So the total weight of TE materials is about 4.2×1.22×0.15×2÷1000=0.0015 g.

2.1816×10⁻⁵×36+0.0624×36=2.25 g.

*h2 Calculated data: Ref 8 presents the output power of per couple is 1.01 μW and the power density is 0.62 W/m². There are 21 TE couples. So the area of TEG is $1.01 \times 21 \div (0.62 \times 10^6) \times 10^4 = 0.34$ cm².

*h3 Estimated data: Ref 8 presents the output power of per couple is 1.01 μW. There are 21 TE couples. So the output power of the TEG is $1.01 \times 21 = 21.21$ μW.

*i1 Estimated data: The length of the whole polyester TE yarn is estimated to be 3.58 m according to the Fig. 3. The density of the polyester TE yarn is assumed to be 14.7625 g/km. So the total weight of TE yarn is about $3.58 \times 14.7625 \div 1000 = 0.053$ g.

*j1 Estimated data: The data is obtained from Fig. 5b in Ref 9.

*j2 Estimated data: The length of the whole CNT TE yarn is estimated to be 8.38 cm according to the Fig. 3. The diameter of the CNT yarn is 40 μm. The density of the CNT yarn is assumed to be 1.2 g/cm³. So the total weight of the TE material is about $\pi \times (2 \times 10^{-3})^2 \times 8.38 \times 1.2 = 1.3 \times 10^{-4}$ g.

**Supplementary References**